\documentclass[12pt]{article}

\ifx\pdfoutput\undefined
\usepackage[dvips,bookmarks]{hyperref}
\else
\usepackage{hyperref}
\fi
\hypersetup{colorlinks=false,bookmarksopen,bookmarksnumbered,citecolor=blue,
   pdfstartview=FitH}

\usepackage{latexsym}
\usepackage{amssymb,amsfonts,amsmath}
\usepackage{graphicx} 
\usepackage{indentfirst}
\usepackage{bbm}
\usepackage{amssymb}
\usepackage{verbatim}
\usepackage{amsmath, amsthm,amssymb}
\usepackage{mathrsfs}
\usepackage{hyperref}
\usepackage{amsfonts}
\usepackage{dsfont}
\usepackage{slashed, tensor}
\usepackage{booktabs}
\usepackage{graphicx}
\usepackage{mathrsfs}

\parskip=\medskipamount

\usepackage[mathscr]{euscript} 

\newcommand {\cA}{{\cal A}}
\newcommand {\cB}{{\cal B}}

\newcommand {\cD}{{\cal D}}
\newcommand {\cE}{{\cal E}}

\newcommand {\cM}{{\cal M}}
\newcommand {\cN}{{\cal N}}

\newcommand {\cP}{{\cal P}}

\newcommand {\cZ}{{\cal Z}}

\def\a{\alpha}
\def\b{\beta}

\def\d{\delta}
\def\e{\epsilon}

\def\g{\gamma}

\def\j{\psi}
\def\k{\kappa}
\def\l{\lambda}
\def\m{\mu}

\def\o{\omega}
\def\p{\pi}
\def\q{\theta}
\def\r{\rho}
\def\s{\sigma}

\def\x{\xi}

\def\D{\Delta}
\def\F{\Phi}
\def\J{\Psi}
\def\L{\Lambda}
\def\O{\Omega}

\def\Q{\Theta}

\def\ri{{\rm i}}
\def\re{{\rm e}}

\newcommand{\gd}{{\dot\g}}

\newcommand{\ad}{{\dot{\alpha}}}
\newcommand{\bd}{{\dot{\beta}}}

\newcommand{\sSU}{\mathsf{SU}}
\newcommand{\sSL}{\mathsf{SL}}

\newcommand{\sSO}{\mathsf{SO}}
\newcommand{\sU}{\mathsf{U}}

\newcommand{\sOSp}{\mathsf{OSp}}

\newcommand{\1}{{\underline{1}}}
\newcommand{\2}{{\underline{2}}}

\newcommand{\ve}{\varepsilon}
\newcommand{\cDB}{{\bar\cD}}
\newcommand{\DB}{\bar{D}}

\newcommand{\pa}{\partial}
\newcommand{\hf}{\frac12}

\newcommand{\vf}{\varphi}

\newcommand{\be}{\begin{equation}}
\newcommand{\ee}{\end{equation}}
\newcommand{\bea}{\begin{eqnarray}}
\newcommand{\eea}{\end{eqnarray}}
\newcommand{\non}{\nonumber}
\newcommand{\ba}{\begin{array}}
\newcommand{\ea}{\end{array}}

\newcommand{\bm}[1]{\mbox{\boldmath$#1$}}

\def\double #1{#1{\hbox{\kern-2pt $#1$}}}

\newcommand{\ts}{{\tilde{\s}}}

\newcommand{\bsubeq}{\begin{subequations}}
\newcommand{\esubeq}{\end{subequations}}

\newcommand{\rd}{\mathrm d}
\numberwithin{equation}{section}  

\newcommand{\mub}{{{\bar{\mu}}}}

\newcommand{\qb}{{\bar{\theta}}}

\newcommand{\lb}{\bar{\l}}
\newcommand{\xb}{\bar{\x}}

\begin{document}

\begin{titlepage}
\begin{flushright}
December, 2015 \\
\end{flushright}
\vspace{5mm}

\begin{center}
{\Large \bf 
Nilpotent chiral superfield in $\cN=2$ supergravity \\
and partial rigid supersymmetry breaking}
\\ 
\end{center}

\begin{center}

{\bf
Sergei M. Kuzenko${}^{a}$ and
Gabriele Tartaglino-Mazzucchelli${}^{b}$
} \\
\vspace{5mm}

\footnotesize{
${}^{a}${\it School of Physics M013, The University of Western Australia\\
35 Stirling Highway, Crawley W.A. 6009, Australia}}  
~\\
\vspace{2mm}
\footnotesize{
${}^{b}${\it Instituut voor Theoretische Fysica, KU Leuven,\\
Celestijnenlaan 200D, B-3001 Leuven, Belgium}
}
\vspace{2mm}
~\\
\texttt{sergei.kuzenko@uwa.edu.au,
Gabriele.Tartaglino-Mazzucchelli@fys.kuleuven.be}\\
\vspace{2mm}

\end{center}

\begin{abstract}
\baselineskip=14pt
In the framework of $\cN=2$ conformal supergravity in four dimensions, 
we introduce a nilpotent chiral superfield suitable for the description of
partial supersymmetry breaking in maximally supersymmetric spacetimes. 
As an application, we construct Maxwell-Goldstone multiplet actions
for partial $\cN=2 \to \cN=1$ supersymmetry breaking on
${\mathbb R} \times S^3$, ${\rm AdS}_3 \times S^1$
(or its covering ${\rm AdS}_3 \times {\mathbb R}$), 
and a pp-wave spacetime. In each of these cases, 
the action coincides with a unique curved-superspace extension of the $\cN=1$ 
supersymmetric Born-Infeld action, which is 
singled out by the requirement of $\sU(1)$ duality invariance. 
\end{abstract}

\vfill

\vfill
\end{titlepage}

\newpage
\renewcommand{\thefootnote}{\arabic{footnote}}
\setcounter{footnote}{0}

\tableofcontents



\allowdisplaybreaks

\section{Introduction}

Inspired by the work of Antoniadis, Partouche and Taylor \cite{APT},
Bagger and Galperin \cite{BG} constructed 
the Goldstone-Maxwell multiplet model for partially broken $\cN=2$ 
Poincar\'e supersymmetry in four spacetime dimensions (4D). 
Their model proved to coincide with the $\cN=1$ supersymmetric Born-Infeld 
action \cite{DP,CF}. Two years later, Ro\v{c}ek and Tseytlin \cite{RT} 
re-derived the model of \cite{BG} using $\cN=2$ superfields, 
building on the earlier formulation due to Ro\v{c}ek \cite{Rocek} 
for the Volkov-Akulov Goldstino model \cite{VA}  
in terms of a nilpotent $\cN=1$ chiral superfield.\footnote{The same nilpotent chiral 
superfield was independently  introduced, a few months later, 
by Ivanov and Kapustnikov 
\cite{IK2}  as a simple application of the
general relationship between linear and nonlinear realisations of supersymmetry
established in their earlier work \cite{IK1}.} 

The $\cN=2$ Minkowski superspace is one of many maximally supersymmetric 
backgrounds in 4D $\cN=2$ off-shell supergravity. Such superspaces were
 classified in \cite{BIL} building on the earlier analysis \cite{KNT-M}
 of maximally supersymmetric 
backgrounds in 5D $\cN=1$ off-shell supergravity.
The construction in \cite{RT} is down-to-earth in the sense that 
it is specifically designed to describe the partial breaking of $\cN=2$
Poincar\'e supersymmetry. Here we present a theoretical scheme
which is suitable for the  description of  partial supersymmetry breaking in 
curved maximally supersymmetric 
backgrounds in 4D $\cN=2$ off-shell supergravity.
As an application of this scheme, we construct Maxwell-Goldstone multiplet 
actions for partial $\cN=2 \to \cN=1$ supersymmetry breaking on
${\mathbb R} \times S^3$,  
${\rm AdS}_3 \times S^1$
(or its covering
${\rm AdS}_3 \times {\mathbb R}$), 
and a pp-wave.  

This paper is organised as follows. In section 2 
we introduce a nilpotent chiral superfield coupled to $\cN=2$ conformal supergravity. 
In section 3 we explain how such a superfield can be used to construct 
a model for partially broken supersymmetry for certain maximally supersymmetric 
backgrounds of $\cN=2$ supergravity. The formalism developed is applied in section 4
to re-derive the Ro\v{c}ek-Tseytlin construction. In section 
5 we construct   Maxwell-Goldstone multiplet actions
for partial $\cN=2 \to \cN=1$ supersymmetry breaking on
${\mathbb R} \times S^3$,  ${\rm AdS}_3 \times S^1$
(or its covering ${\rm AdS}_3 \times {\mathbb R}$), 
and a pp-wave. 
Concluding comments are given in section 6.
The main body of the paper is accompanied by three technical appendices. 
In Appendices A and B, we present group-theoretic formulations 
for four-dimensional $\cN=1$ and $\cN=2$ superspaces over
 $\sU(2)=(S^1 \times S^3)/{\mathbb Z}_2$. 
The maximally $\cN=2$ supersymmetric background over ${\mathbb R} \times S^3$, 
which is used in section 5, is the universal covering space of the $\cN=2$
superspace over $(S^1 \times S^3)/{\mathbb Z}_2$.
Appendix A also contains the  group-theoretic description of $\cN=1$ superspace over 
$\sU(1,1)=({\rm AdS}_3 \times S^1)/{\mathbb Z}_2$.
Appendix C is devoted to the discussion of a unique feature of the anti-de Sitter supersymmetry
that distinguishes AdS${}_4$ from the other maximally supersymmetric four-dimensional backgrounds.

\section{Nilpotent chiral superfield in $\cN=2$ supergravity}\label{section2}

In the framework of four-dimensional $\cN=2$ conformal supergravity\footnote{In
this paper, we use Howe's superspace formulation \cite{Howe}  for $\cN=2$ 
conformal supergravity and follow the supergravity notation and conventions of 
\cite{KLRT-M2}. In particular, the superspace covariant derivatives are denoted 
${\cD}_{\cA} =({ \cD}_{a}, { \cD}_{{\a}}^i, { \cDB}^\ad_i)$.
We make use of the second-order differential operators
$\cD^{ij}:=\cD^{\a(i}\cD_\a^{j)}$, $\cDB^{ij}:=\cDB_\ad^{(i}\cDB^{\ad j)}$.
The $\sSU(2)$ triplet ${S}^{ij}= S^{ji}$ and its conjugate 
$\bar{S}_{ij} = \overline{S^{ij}}$ stand for certain components
of the superspace torsion tensor.}
we introduce a nilpotent chiral superfield constrained by 
\begin{subequations}\label{Z}
\bea
\bar \cD^i_\ad \cZ &=&0~, \label{Za} \\
\big(\cD^{ij}+4S^{ij}\big)\cZ
- \big(\cDB^{ij}+ 4\bar{S}^{ij}\big)\bar{\cZ} &=&4  \ri \,G^{ij} ~, \label{Zb}\\
\cZ^2 &=&0~, \label{Zac}
\eea
\end{subequations}
where $G^{ij}$  is a linear multiplet constrained by $G^{ij} G_{ij} \neq 0$.
One may interpret $G^{ij}$ as the field strength of a  tensor multiplet. 
The constraints \eqref{Za}--\eqref{Zac} are invariant under the $\cN=2$ super-Weyl 
transformations \cite{Howe,KLRT-M2} if $\cZ$ is considered to be a primary superfield of dimension 1.

A chiral superfield constrained by 
 \eqref{Zb} was considered in \cite{K-superWeyl}
 in the context of the dilaton effective action in $\cN=2$ supergravity.
In the super-Poincar\'e case, chiral superfields obeying the constraint 
\eqref{Zb} with a constant $G^{ij} $ naturally originate in the framework
of partial $\cN=2 \to \cN=1 $ supersymmetry breaking \cite{APT,IZ1,IZ2}.

We recall that the $\cN=2$ tensor multiplet is  described in curved superspace by
its gauge invariant field strength $G^{ij}$  which is 
a linear multiplet. The latter is 
defined to be a  real ${\sSU}(2)$ triplet (that is, 
$G^{ij}=G^{ji}$ and ${\bar G}_{ij}:=\overline{G^{ij}} = G_{ij}$)
subject to the covariant constraints  \cite{BS,SSW}
\bea
\cD^{(i}_\a G^{jk)} =  {\bar \cD}^{(i}_\ad G^{jk)} = 0~.
\label{1.2}
\eea
These constraints are solved in terms of a chiral
prepotential $\Psi$ \cite{HST,GS82,Siegel83,Muller86} via
\begin{align}
\label{eq_Gprepotential}
G^{ij} = \frac{1}{4}\Big( \cD^{ij} +4{S}^{ij}\Big) \Psi
+\frac{1}{4}\Big( \cDB^{ij} +4\bar{S}^{ij}\Big){\bar \Psi}~, \qquad
{\bar \cD}^i_\ad \J=0~,
\end{align}
which is invariant under Abelian gauge transformations
\bea
\d_\L \Psi = \ri \Lambda~,
\label{TMGT}
\eea
with $\Lambda$ a reduced chiral superfield,
\begin{subequations}
\bea
\bar \cD^i_\ad \L &=&0~,\\
\big(\cD^{ij}+4S^{ij}\big)\L
- \big(\cDB^{ij}+ 4\bar{S}^{ij}\big)\bar{\L} &=& 0~.
\eea
\end{subequations}
We recall that the field strength of an Abelian vector multiplet is a reduced chiral superfield \cite{GSW}.

The constraints on $\L$ can be solved in terms of 
the Mezincescu prepotential \cite{Mezincescu} (see also \cite{HST}),  $U_{ij}=U_{ji}$,
which is an unconstrained real $\sSU(2)$ triplet. 
The curved-superspace solution is \cite{BK11}
\begin{align}
\L = \frac{1}{4}\bar\Delta \Big({\cD}^{ij} + 4 S^{ij}\Big) U_{ij}~.
\end{align}
Here   $\bar{\D}$ denotes the chiral projection operator \cite{KT-M09,Muller}
\bea
\bar{\D}
&=&\frac{1}{96} \Big((\cDB^{ij}+16\bar{S}^{ij})\cDB_{ij}
-(\cDB^{\ad\bd}-16\bar{Y}^{\ad\bd})\cDB_{\ad\bd} \Big)
\non\\
&=&\frac{1}{96} \Big(\cDB_{ij}(\cDB^{ij}+16\bar{S}^{ij})
-\cDB_{\ad\bd}(\cDB^{\ad\bd}-16\bar{Y}^{\ad\bd}) \Big)~,
\label{chiral-pr}
\eea
with $\cDB^{\ad\bd}:=\cDB^{(\ad}_k\cDB^{\bd)k}$.
Its main properties can be formulated using 
a super-Weyl inert scalar $V$. It holds that
\begin{subequations} 
\bea
{\bar \cD}^{\ad}_i \bar{\D} V &=&0~, \\
\d_\s V = 0 \quad \Longrightarrow \quad 
\d_\s \bar \D V &=& 2\s \bar \D V~,  \label{2.5b}\\
\int \rd^4 x \,{\rm d}^4\q\,{\rm d}^4{\bar \q}\,E\, V
&=& \int {\rm d}^4x \,{\rm d}^4 \q \, \cE \, \bar{\D} V ~,
\label{chiralproj1} 
\eea
\end{subequations}
where the real unconstrained parameter $\s$ corresponds to the super-Weyl transformations \cite{KLRT-M2}.\footnote{The parameter $\s$ was denoted $2U$ 
in \cite{KLRT-M2}.}
Here $E$ and $\cE$ denote the full superspace and chiral densities, respectively. 

The constraints \eqref{Za} and  \eqref{Zb} define a deformed 
reduced chiral superfield. These constraints may be re-cast in the language 
of superforms as $\rd F = H$, 
where $F$ is a two-form and $H$ is the  three-form field strength, $\rd H=0$,
describing the tensor multiplet \cite{Muller}, 
see also \cite{BN}.\footnote{We are grateful to Joseph Novak for this observation.}
Switching $H$ off,  $H=0$, turns $F$ into the two-form field strength of the vector multiplet. 

The constraint \eqref{Zb} naturally  originates as follows. 
Consider the model for a massive improved tensor multiplet 
coupled to $\cN=2$ conformal supergravity \cite{K-tensor,K-08}. 
The action of this model in the form given in \cite{BK11} is 
\bea
S_{\rm tensor}  = - \int \rd^4 x \,{\rm d}^4\q \, \cE \, \Big\{
\J {\mathbb W} + \frac{1}{4}\m(\m+\ri e)  \J^2 \Big\} +{\rm c.c.} ~, 
\label{2.11}
\eea
where  $\m$ and $e$ are real parameters,  with $\m \neq 0$
(the tensor multiplet mass can be shown to be $m=\sqrt{\m^2 + e^2}$).
The kinetic term involves the composite \cite{deWPV}
\bea
   \mathbb W := -\frac{G}{8} (\bar \cD_{ij} + 4 \bar S_{ij}) \left(\frac{G^{ij}}{G^2} \right) ~, 
   \label{2.100}
\eea
which proves to be a reduced chiral superfield.\footnote{The superfield \eqref{2.100} is 
one of the simplest applications of the powerful approach 
to generate composite reduced chiral multiplets 
which was presented in \cite{BK11}.}
For $m=0$ the above action describes the improved tensor multiplet \cite{deWPV}.
We introduce a St\"uckelberg-type extension of the model
\bea
\widetilde{S}_{\rm tensor}  = - \int \rd^4 x \,{\rm d}^4\q \, \cE \, \Big\{
\J {\mathbb W} + \frac{1}{4}\m(\m+\ri e)  (\J - \ri W)^2 \Big\} +{\rm c.c.} ~, 
\label{2.13}
\eea
where $W$ is the field strength of a vector multiplet. 
The action is invariant under the gauge transformation \eqref{TMGT}
accompanied by 
\bea
\d_\L W = \L~.
\label{2.14}
\eea
The original action \eqref{2.11} is obtained from \eqref{2.13}
by choosing a gauge $W=0$. 
Now one can see that the superfield $\cZ := W +\ri \J$ obeys the constraint 
\eqref{Zb}. 

It is well known that the functional 
\bea
\ri  \int \rd^4 x \,{\rm d}^4\q \, \cE \,  W^2  +{\rm c.c.} 
\eea
is a total derivative. Since the mass term in \eqref{2.13} is invariant under the gauge transformation \eqref{TMGT} and \eqref{2.14}, it follows that,   
given a chiral superfield $\cZ$ constrained by \eqref{Zb},  the  functional 
\bea
I=  \int {\rm d}^4x \,{\rm d}^4 \q \, \cE \, \Big\{ \cZ \J - \frac{\ri}{2}  \J^2 \Big\} 
~+~{\rm c.c.}
\eea
is invariant under the gauge transformation \eqref{TMGT},
$\d_\L I =0$.

The constraints  \eqref{Za}--\eqref{Zac} imply that, for certain 
supergravity backgrounds,  
the degrees of freedom described by the $\cN=2$ chiral superfield $\cZ$ 
are in a one-to-one correspondence with those of an Abelian $\cN=1$ vector multiplet. 
The specific feature of such $\cN=2$ supergrvaity backgrounds is that 
they possess an $\cN=1$ subspace of the full $\cN=2$ superspace. 
This property is not universal. In particular, there exist maximally 
$\cN=2$ supersymmetric backgrounds with no admissible truncation to $\cN=1$ 
\cite{BIL}.


\section{Maximally $\cN=2$ supersymmetric backgrounds and partial supersymmetry breaking}
\label{section_3}

So far we have discussed an arbitrary supergravity background. 
Now we restrict our consideration to a maximally supersymmetric background 
${\mathbb M}^{4|8}$
with the property that the chiral prepotential $\J $ for $G^{ij}$ may be chosen such that the following two conditions hold. Firstly, the complex linear multiplet
\bea
G_+^{ij} := \frac{1}{4}\Big( \cD^{ij} +4{S}^{ij}\Big) \Psi
\label{3.1}
\eea
is covariantly constant and null, 
\bea
\cD_\cA G_+^{ij} &=&0 ~, \label{3.2}\\
G_+^{ij}G_{+ij} &=&0~.\label{3.3}
\eea
Secondly, the prepotential $\J$ may be chosen to be nilpotent, 
\bea
\J^2=0~.
\label{3.4}
\eea
 The null condition for $G_+^{ij} $ means that 
$G_+^{ij} = q^i q^j$, for some isospinor $q^i$. 
It follows that $G^{ij} =G^{ij}_+ + G^{ij}_-$  is covariantly constant, 
\bea
\cD_\cA G^{ij}=0~,
\eea
where we have denoted 
$G_-^{ij} := \frac{1}{4}\Big( \bar \cD^{ij} +4\bar{S}^{ij}\Big) \bar \Psi$.

We are going to show that the following functional 
\bea
I=  \int {\rm d}^4x \,{\rm d}^4 \q \, \cE \, \J \cZ  
\eea
is supersymmetric. 
Here $\cZ$ is the nilpotent chiral superfield \eqref{Z}, which is assumed to  
be a composite of the dynamical fields.
The  complex linear multiplet \eqref{3.1}
 and its chiral prepotential $\J$ are  background fields
 associated with the background superspace ${\mathbb M}^{4|8}$. 
 Since the covariant derivatives $\cD_\cA$ are invariant under the
isometry transformations of ${\mathbb M}^{4|8}$, the fields 
$G^{ij}_+$ and $\J$ do not change under such transformations. 
Let $\x$ be a Killing supervector field for ${\mathbb M}^{4|8}$ 
(see section 6.4 of \cite{BK} and \cite{K15Corfu} for general discussions).
 Then 
\bea
\d_\x I=  \int {\rm d}^4x \,{\rm d}^4 \q \, \cE \,  \J \d_\x \cZ 
= -  \int {\rm d}^4x \,{\rm d}^4 \q \, \cE \,  \cZ \d_\x \J ~.
\eea
We introduce a reduced chiral superfield $W$ by
\bea
\cZ = W +  \ri \J~, \qquad
W = \frac{1}{4}\bar\Delta \Big({\cD}^{ij} + 4 S^{ij}\Big) U_{ij}~,
\eea
where $U_{ij}$ is the Mezincescu prepotential for the reduced chiral superfield $W$. 
Since $\J \d_\x \J =0$, 
we have 
\bea
\d_\x I = -  \int {\rm d}^4x \,{\rm d}^4 \q \, \cE \,  \cZ \d_\x \J 
&=& -  \int {\rm d}^4x \,{\rm d}^4 \q \, \cE \,  W \d_\x \J \non \\
&=& - \frac{1}{4}  \int {\rm d}^4x \,{\rm d}^4 \q \, \rd^4 \bar \q\, E \,
U_{ij} \Big({\cD}^{ij} + 4 S^{ij}\Big) \d_\x \J \non \\
&=& - \frac{1}{4}  \int {\rm d}^4x \,{\rm d}^4 \q \, \rd^4 \bar \q\, E \,
U_{ij} \d_\x \Big({\cD}^{ij} + 4 S^{ij}\Big)  \J \non\\
&=& -   \int {\rm d}^4x \,{\rm d}^4 \q \, \rd^4 \bar \q\, E \,
U_{ij} \d_\x G_+^{ij} =0~.
\eea

In the next two sections, it will be shown that the action
\bea
S=  -\frac{\ri}{4} \int {\rm d}^4x \,{\rm d}^4 \q \, \cE \, \J \cZ  +{\rm c.c.}
\label{action_0}
\eea
describes the Maxwell-Goldstone multiplet 
for partial $\cN=2 \to \cN=1$ supersymmetry breaking
on  the maximally supersymmetric backgrounds specified. 

The above derivation does not use the null condition \eqref{3.3}.
The latter is introduced for the $\cN=2$ superspace ${\mathbb M}^{4|8}$
to possess an $\cN=1$ subspace. 


\section{Example: The super-Poincar\'e case}

The simplest maximally supersymmetric background is $\cN=2$ 
Minkowski superspace.
In this superspace, 
every constant  real $\sSU(2)$ triplet  $G^{ij}$ is covariantly constant,  
\bea
D_\cA G^{ij}=0
~,
\eea
where $D_\cA = (\pa_a, D_\a^i , \bar D^\ad_i) $ are the flat superspace covariant derivatives. 
Let $\Psi$ be a chiral prepotential for $G^{ij}$, $\DB^i_\ad \J=0$.
We represent 
\bea
G^{ij} =G_+^{ij}+{G}_-^{ij}~, \qquad
G^{ij}_+ =  \frac{1}{4}D^{ij} \Psi~, 
\qquad
G^{ij}_-=\frac{1}{4}\DB^{ij}{\bar \Psi}~.
\eea
It is always possible to choose the prepotential $\J$  such that
the following properties hold:
\bea
\J^2=0
~,\qquad
D_\cA G_+^{ij}=0
~,\qquad
G_+^{ij}G_{+ij}=0
~.
\eea

In $\cN=2$ Minkowski superspace, the constraints \eqref{Za}--\eqref{Zac} turn into
\begin{subequations}
\label{constr_Z_0}
\bea
\DB^i_\ad \cZ &=&0~, \\
D^{ij}\cZ
- \DB^{ij}\bar{\cZ} &=&4 \ri \,G^{ij} ~,\\
\cZ^2 &=&0
~.
\eea
\end{subequations}
The action \eqref{action_0}  becomes
\bea
S=
-\frac{\ri}{4}\int {\rm d}^4x \,{\rm d}^4 \q \, \cZ \J 
~+~{\rm c.c.}
\label{action_1}
\eea
Since $G_+^{ij}$ is constant, 
it is invariant under the $\cN=2$ supersymmetry transformations.
In accordance with the analysis given in the previous section, 
the action is $\cN=2$ supersymmetric.

For the Grassmann coordinates $\q^\a_i$ and $\bar \q_\ad^j$
of $\cN=2$ Minkowski superspace, 
as well as for the spinor covariant derivatives $D_\a^i $ and $ \bar D^\ad_i$, 
it is useful to label the values of their $R$-symmetry  indices as $i, j = \1, \2$.
Without loss of generality we can choose
\bea
G_{+}^{ij}=-\ri\d^i_\2\d^j_\2~,~~~~~~
\Psi=\ri
\q^\a_{\2}\q_{\a\2}
~.
\eea
We can now reproduce the results of \cite{BG} from the  $\cN=2$ 
setup described. In order to solve the constraints \eqref{constr_Z_0}, 
it is useful to carry out a reduction to $\cN=1$ Minkowski superspace. 

Given a superfield  $U(x,\q_i,\qb^i)$ on $\cN=2$ 
Minkowski superspace, we introduce
its bar-projection
\bea
U|:=U(x,\q_i,\qb^i)|_{\q_\2={\bar \q}^\2=0}~,
\eea
which is a superfield on $\cN=1$ 
Minkowski superspace with the Grassmann coordinates $\q^\a = \q^\a_\1$ 
and $\bar \q_\ad = \bar \q_\ad^\1$ and the spinor covariant derivatives $D_\a =D_\a^\1$
and $\bar D^\ad = \bar D^\ad_\1$.
The background superfield $\Psi$ is characterised by the properties 
\bea
\Psi|
=0~,\qquad D_\a^\2\Psi| =0~.
\eea

Since 
$\cZ^2=0$,
the constraints \eqref{constr_Z_0}  imply
\bea
(D^{\a\2}\cZ)D_\a^\2\cZ
+\cZ\DB_{\1\1}\bar{\cZ} 
+4 \cZ
=0
~.
\eea
Taking the bar-projection of this constraint gives
\bea
X
+\frac{1}{4}X\DB^2\bar{X}
&=&
W^2~,~~~~~~
W^2:=W^\a W_\a
~,
\label{BG}
\eea
where we have introduced the $\cN=1$ components of $\cZ$:
\bsubeq
\bea
&&X:=\cZ|
~,\qquad
W_\a
:=
-\frac{\ri}{2}D_\a^\2 \cZ|
~.
\label{4.11a}
\eea
These superfields satisfy the constraints
\bea
\DB_\ad X&=&0~,\qquad
\DB_\ad W_\a=0
~,\qquad
D^\a W_\a=\DB_\ad \bar{W}^\ad
~.
\eea
\esubeq
The constraints on  $W_\a$ tell us that it can be interpreted 
as  the field strength of an Abelian $\cN=1$ vector multiplet.
The constraint \eqref{BG} is equivalent to the Bagger-Galperin constraint \cite{BG}.
Its general solution is
\bsubeq\label{constrX-000}
\bea
X&=&
W^2
-\frac{1}{2}\DB^2\frac{W^2\bar{W}^2}{
\Big(\,
1
+\hf A
+\sqrt{1+A+\frac{1}{4}B^2}
\,\Big)
}
~,
\\
A&=&\frac{1}{2}\big(D^2W^2+\DB^2\bar{W^2}\big)
~,~~~~~~
B=\frac{1}{2}\big(D^2W^2-\DB^2\bar{W^2}\big)
~.
\eea
\esubeq
Upon reduction to $\cN=1$ superspace, 
the action \eqref{action_1}  becomes
\bea
I=\frac{1}{4}\int {\rm d}^4x \,{\rm d}^2 \q \, X 
+\frac{1}{4}\int {\rm d}^4x \,{\rm d}^2 \qb \,\bar X
~.
\label{DBI_action}
\eea
This is the $\cN=1$ supersymmetric Born-Infeld  action.
Being manifestly $\cN=1$ supersymmetric, 
the action is also invariant under the second nonlinearly realised 
supersymmetry transformation \cite{BG}
\bea
\d_\e W_\a
=
\e_\a
+\frac{1}{4}\e_\a\DB^2\bar{X}
+\ri\bar{\e}^\bd\pa_{\a\bd}X \quad \Longrightarrow \quad 
\d_\e X=2\e^\a W_\a
~.
\label{transf_0}
\eea
For completeness, we  re-derive this result.

Let $U$ be a scalar superfield on $\cN=2$ Minkowski superspace. 
Its isometry transformation is 
\bea
\d_\x U = - \x U~, 
\eea
where 
\be
\x = {\overline \x} = \x^\cA D_\cA = \x^a \pa_a + \x^\a_i D^i_\a
+ {\bar \x}_\ad^i  {\bar D}^\ad_i
\ee   
is a Killing supervector field of Minkowski superspace,\footnote{It follows from 
\eqref{C.3} that $\x^\a_i$ is chiral, $\bar D_\bd^j \x^\a_i =0$.} 
\bea
\x^\a_i = -\frac{\rm i}{8} {\bar D}_{\bd i} \x^{\bd \a}~, \qquad 
  D^i_{(\a} \x_{\b )\bd} = {\bar D}_{i(\ad} \x_{\b \bd )}=0~,
  \qquad D_\a^i \x^\a_i =0~.
\label{C.3} 
\eea

The Killing supervector field generating the supersymmetry transformation is
characterised by the components 
\bea
\x^a =2\ri (\q_i\s^a \bar \e^i - \e_i \s^a \bar \q^i)~, \qquad 
\x^\a_i = \e^\a_i = {\rm const}~.
\eea
Applying this transformation to $\cZ$ gives 
$\d_\x \cZ  = - (\x^a \pa_a +\x^\a_i D^i_\a)\cZ$.
We now consider only the second supersymmetry transformation 
by choosing  $\e^\a_\1=0$ and $\e^\a_\2 =\e^\a$. 
It acts on the $\cN=1$ superfields \eqref{4.11a} as follows
\begin{subequations}\label{C.5}
\bea
\d_\e X &=&  \d_\x \cZ|=-(\x \cZ)| = - \e^\a (D^\2_\a \cZ)| = -2\ri \e^\a W_\a~, \\
\d_\e W_\a &=&- \frac{\ri}{2}( D_\a^\2 \d_\x \cZ)|
= -\ri\e_\a 
-\frac{\ri}{4}\e_\a\DB^2\bar{X}
-\bar{\e}^\bd \pa_{\a\bd} X~,
\eea
\end{subequations}
where we have made use of the constraints obeyed by $\cZ$ and $X$. 
The supersymmetry transformation \eqref{transf_0} follows from \eqref{C.5}
upon a rescaling of $\e^\a$.


\section{Maxwell-Goldstone multiplet for partially broken rigid supersymmetry 
in curved space}

We turn to applying the theoretical framework of section \ref{section_3} 
to maximally supersymmetric curved backgrounds in $\cN=2$ supergravity.

\subsection{Curved $\cN=2$ superspace backgrounds}

We consider a maximally supersymmetric background 
${\mathbb M}^{4|8}$ described by 
the following algebra of covariant derivatives\footnote{Here $M_{ab}$, $J^{kl}$ and $Y$ are 
the Lorentz, $\sSU(2)$ and $\sU(1)$ generators, respectively, defined as in \cite{KLRT-M2}.}
\bsubeq \label{5.1}
\bea
\{\cD_\a^i,\cD_\b^j\}&=&\{\cDB^\ad_i,\cDB^\bd_j\}=
0~,\\
\{\cD_\a^i,\cDB^\bd_j\}&=&
-2\ri\d^i_j\cD_\a{}^\bd
+4\ri G^{\g\bd}{}^{i}{}_{j}M_{\a\g}
+4\ri G_{\a\gd}{}^{i}{}_{j}\bar{M}^{\gd\bd}
\non\\
&&
-4\ri\d^i_jG_{\a}{}^\bd{}^{kl}J_{kl}
-2\ri G_{\a}{}^{\bd}{}^i{}_jY
~,~~~~~~~~~
\\
{[}\cD_a,\cD_\b^j{]}&=&
(\ts_a)^{\ad\g}G_{\b\ad}{}^{j}{}_k\cD_\g^k
~,~~~~~~
{[}\cD_a,\cDB_{\bd j}{]}=
-(\ts_a)^{\gd \a}G_{\a\bd}{}_{j}{}^{k}\cDB_{\gd k}
~,
\eea
where the torsion tensor $G_{a}^{ij}$ is annihilated by the spinor 
covariant derivatives, 
\bea
\cD_\a^i G_b^{jk} = 0~, \qquad \bar \cD_\ad^i G_b^{jk} =0~. 
\label{5.2}
\eea
\esubeq
This algebra is obtained from that corresponding to $\cN=2$ conformal
supergravity, and given by eq. (2.8) in \cite{KLRT-M2}, by (i) switching off 
the components $S^{ij}$, $Y_{\a\b}$, $W_{\a\b}$ and $G_{\a\ad}$ 
of the torsion tensor; and (ii) imposing  \eqref{5.2}. 
The constraints \eqref{5.2} are required by the theorem 
\cite{KNT-M} that all fermionic components of the superspace torsion tensor 
must vanish in maximally supersymmetric backgrounds.  

In complete analogy with the 5D case \cite{KNT-M}, 
 the constraints \eqref{5.2} imply the following integrability condition 
\bea
G_a{}^{k(i} G_b{}^{j)}{}_k=0~.
\label{5.3}
\eea
As shown in \cite{KNT-M}, the general solution of the conditions
\eqref{5.2} and \eqref{5.3} is 
\bea
&&G_b{}^{kl}= -\frac{1}{4}g_b s^{kl}
~,\qquad
\cD^i_\a g_b=0~,\quad \bar \cD^i_\ad g_b =0~, \qquad
\cD_A s^{kl}=0~,
\eea
for some real vector $g_b$ and real $\sSU(2)$ triplet $s^{kl}$.
The latter may be normalised as 
\bea
s^{ij}s_{ij}=2~.
\eea
Since $g^2 = g^a g_a$ is constant, $\cD_A g^2 =0$,
there are in fact three different superspaces
described by the above algebra: (i) if $g_a$ is time-like, $g^2<0$, 
the bosonic body of ${\mathbb M}^{4|8}$ is ${\mathbb R}\times S^3$; 
(ii) if $g_a$ is space-like, $g^2 >0$, the bosonic body of ${\mathbb M}^{4|8}$ is 
${\rm AdS}_3 \times {\mathbb R}$; (iii) in the null case, $g^2=0$, 
the spacetime geometry is a pp-wave. 
We will denote these superspaces as ${\mathbb M}^{4|8}_{T}$,
 ${\mathbb M}^{4|8}_{S}$ and ${\mathbb M}^{4|8}_{N}$, respectively.  
These backgrounds were  constructed in \cite{BIL}, and 
they have 5D cousins \cite{KNT-M}.

In order to get some more insight into the structure of 
the superspace geometry \eqref{5.1}, 
a specific value  of $g^2$ has to be fixed.  It suffices to consider the superspace
 ${\mathbb M}^{4|8}_{T}$, since the other two cases may be treated similarly. 
As a  supermanifold, ${\mathbb M}^{4|8}_{T}$ is the universal covering of 
the 4D $\cN=2$ superspace introduced in Appendix B. 

In the case $g^2<0$, 
it is possible to choose a Lorentz and $\sSU(2)_R$ gauge such that 
\bea
g_a=(g,0,0,0)~,\qquad
s_i{}^j
=
-\ri(\s^3)_i{}^j=\ri (-1)^i \d_i^j
~.
\eea
As shown in \cite{BIL}, 
the algebra of covariant derivatives is equivalent to 
\bsubeq
\label{SU21SU21}
\bea
\{\cD_\a^i,\cD_\b^j\}=\{\cDB^\ad_i,\cDB^\bd_j\}&=&
0~,
~~~~~~
\{\cD_\a^i,\cDB^\bd_j\}=
-2\ri\d^i_j(\s^a)_\a{}^\bd\cD_a^{(i)}
~,
\\
{[}\cD_a^{(i)},\cD_\b^j{]}
&=&
\frac{\ri}{2}  \d^{ij}(-1)^{j}(\s_a)_{\b\bd}\,g^{\bd\g}\cD_\g^j
~,
\\
{[}\cD_a^{(i)},\cD_b^{(j)}{]}
&=&
(-1)^{j+1} \d^{ij}\ve_{abc}{}^{d}g^c\cD_d^{(j)}
~,
\eea
\esubeq
where we have introduced  the  ``improved'' vector covariant derivatives 
\bea
\cD_a^{(i)}
:=\cD_a
+\hf g_a s^{kl}J_{kl}
+(-1)^i\Big(
\frac{1}{4}\ve_{abcd}\,g^bM^{cd}
+\ri g_aY
\Big)
~.
\eea
These (anti-)commutation relations correspond to the superalgebra
 $\frak{su}(2|1)\times \frak{su}(2|1)$.

The superspace geometry of  ${\mathbb M}^{4|8}_{T}$
can be described, e.g., in terms of the 
covariant derivatives $\tilde{\cD}_\cA=(\cD_a^{(\1)} ,\cD_\a^i,\cDB^\ad_i)$.
In accordance with \eqref{SU21SU21}, 
the operators 
$(\cD_a^{(\1)}, \cD_\a^\1,\cDB^\ad_\1)$ form a closed algebra isomorphic
to that of the superalgebra $\frak{su}(2|1)$.
 This property means 
that the $\cN=2$ superspace ${\mathbb M}^{4|8}_{T}$ 
possesses an $\cN=1$ subspace  which will be denoted ${\mathbb M}^{4|4}_{T}$.
It turns out that all the conditions \eqref{3.2}--\eqref{3.4}
can be met in the case of  ${\mathbb M}^{4|8}_{T}$.
In particular, this superspace allows the existence of covariantly  
constant complex $\sSU(2)$ triplets $G^{ij}_+$.
Since the graded commutation relations for $\tilde{\cD}_\cA$ involve 
 neither Lorentz nor $\sSU(2)$ curvature tensors,  
 the Lorentz and $\sSU(2)$ connections may be gauged away.
 In such a gauge, every constant complex $\sSU(2)$ triplets $G^{ij}_+$
 is covariantly constant.

Since the superspaces ${\mathbb M}^{4|8}_{T}$,
 ${\mathbb M}^{4|8}_{S}$ and ${\mathbb M}^{4|8}_{N}$
 meet the requirements  \eqref{3.2}--\eqref{3.4}, the formalism 
 of section \ref{section_3} may be used to 
 construct a Maxwell-Goldstone multiplet action
for partial supersymmetry breaking. Instead of implementing the scheme 
directly, we will take a shortcut to constructing such actions 
on the $\cN=1$ subspaces of the superspaces ${\mathbb M}^{4|8}_{T}$,
 ${\mathbb M}^{4|8}_{S}$ and ${\mathbb M}^{4|8}_{N}$.


\subsection{Goldstone multiplet for partially broken supersymmetry}

We consider a maximally supersymmetric background 
${\mathbb M}^{4|4}$ described by
the following algebra of $\cN=1$ covariant derivatives:
\begin{subequations}
\label{RS^3}
\bea
&\{\cD_\a,\cD_\b\}= 0~, \qquad \{\cDB_\ad,\cDB_\bd\}=0~,\qquad
\{\cD_\a,\cDB_\bd\}=-2\ri\cD_{\a\bd}~,
\\
&{[}\cD_\a,\cD_{\b\bd}{]}=\ri\ve_{\a\b}G^\g{}_{\bd}\cD_\g
~,\qquad
{[}\cDB_\ad,\cD_{\b\bd}{]}=-\ri\ve_{\ad\bd}G_\b{}^\gd\cDB_\gd~,
\\
&{[}\cD_{\a\ad},\cD_{\b\bd}{]}=
-\ri\ve_{\ad\bd}G_\b{}^\gd\cD_{\a\gd}
+\ri\ve_{\a\b}G^\g{}_\bd\cD_{\g\ad}
~,
\eea
where the torsion tensor $G_{a}$ is covariantly constant,
\bea
\cD_A G_b = 0~.
\label{5.9}
\eea
\end{subequations}
This is a special case of the superspace geometry for $\cN=1$ old minimal supergravity \cite{GWZ} reviewed in \cite{BK}.
The above algebra is obtained from the supergravity (anti-)commutation relations
(5.5.6) and (5.5.7) in \cite{BK} by (i) switching off the chiral torsion superfields
$R$ and $W_{\a\b\g}$ and their conjugates; and (ii) imposing the condition \eqref{5.9}.

Since $G^2 = G^a G_a $ is constant, the geometry 
\eqref{RS^3} describes  three different superspaces, 
${\mathbb M}^{4|4}_{T}$, ${\mathbb M}^{4|4}_{S}$ and ${\mathbb M}^{4|4}_{N}$, 
which correspond to the choices $G^2<0$, $G^2>0$ and $G^2=0$, 
respectively. These $\cN=1$ superspaces originate as the $\cN=1$ subspaces
of the $\cN=2$ superspaces of ${\mathbb M}^{4|8}_{T}$,
 ${\mathbb M}^{4|8}_{S}$ and ${\mathbb M}^{4|8}_{N}$, respectively, 
 considered in the previous subsection.\footnote{In the time-like case, 
 $G^2<0$, the  graded commutation relations \eqref{RS^3} are obtained from \eqref{SU21SU21}
 by choosing $i, j = \1$ and setting 
$G_a=g_a$.}
We recall that the Lorentzian manifolds supported by these superspaces are 
${\mathbb R}\times S^3$, ${\rm AdS}_3 \times S^1$ or its covering 
${\rm AdS}_3 \times {\mathbb R}$, 
and a pp-wave spacetime, respectively.\footnote{$\cN=1$ supersymmetric theories 
on ${\mathbb R}\times S^3$ were studied in the mid-1980s  by Sen \cite{Sen}.
At  the component level, the maximally $\cN=1$ supersymmetric 
backgrounds in four dimensions were classified by Festuccia and Seiberg \cite{FS}.
Their results were re-derived in \cite{K13} using the superspace formalism
developed in the mid-1990s \cite{BK}.}
As a  supermanifold, ${\mathbb M}^{4|4}_{T}$ is the universal covering of 
the $\cN=1$ superspace $\cM^{4|4}$ 
introduced in section \ref{subsectionA.1}. 
The isometry group of  $\cM^{4|4}$ is $\sSU(2|1) \times \sU(2) $.
As a  supermanifold, ${\mathbb M}^{4|4}_{S}$ is the universal covering of 
the $\cN=1$ superspace $\widetilde{\cM}^{4|4}$ 
introduced in section \ref{subsectionA.2}. The isometry group of
$\widetilde{\cM}^{4|4}$ is $\sSU(1,1|1) \times \sU(2) $.

The superspace ${\mathbb M}^{4|4}$ allows the existence of covariantly constant spinors,
\bea
\cD_A \e_\a =0~.
\label{5.10}
\eea
Such a spinor is constant in a gauge in which the Lorentz connection vanishes.

By analogy with the flat-superspace  case, we consider 
the following $\cN=1$ supersymmetric theory with action
\bea
S=
\frac{1}{4}\int {\rm d}^4x \,{\rm d}^2 \q \, \cE\,X
+{\rm c.c.}~,
\label{curvedAc}
\eea
where the covariantly chiral superfield $X$ is a  unique solution of the constraint
\bea
X+\frac{1}{4}X\cDB^2\bar X=W^2
~.
\label{constrX}
\eea
The superfield $W_\a$ is the chiral field strength of an Abelian vector multiplet
and, together with its complex conjugate $\bar W_\ad$,
it obeys the Bianchi identity
\bea
\cD^\a W_\a=\cDB_\ad \bar W^{\ad}~.
\label{VMBI}
\eea
The explicit solution of the constraint \eqref{constrX}
is a covariantisation  of that described 
in the previous section. It is given, e.g., in  \cite{KMcC}.

The action \eqref{curvedAc} is invariant under a second supersymmetry given by
\bea
\d_\e X=2\e^\a W_\a
~,
\label{eX}
\eea
with the parameter $\e_\a$ being constrained as in  \eqref{5.10}.
Of course, this transformation should be induced by that of $W_\a$. 
The correct supersymmetry transformation of $W_\a$ 
proves to be
\bea
\d_\e W_\a
&=&
\e_\a
+\frac{1}{4}\e_\a\cDB^2\bar{X}
+\ri\bar{\e}^\bd\cD_{\a\bd}X
-\bar{\e}^{\bd}G_{\a\bd}X
~.
\eea
It has the correct flat superspace limit \cite{BG}, compare with \eqref{transf_0},
and respects the Bianchi identity
\eqref{VMBI}, 
\bea
\cD^\a \d_\e W_\a=\cDB_\ad \d_\e \bar W^{\ad}~.
\eea

The dynamical system defined by eqs. \eqref{curvedAc} and \eqref{constrX}
describes the Maxwell-Goldstone multiplet action
for partial $\cN=2 \to \cN=1$ supersymmetry breaking in those curved spacetimes 
which are supported by the superspace geometry \eqref{RS^3}, including 
${\mathbb R}\times S^3$,  ${\rm AdS}_3 \times S^1$ and its covering 
 ${\rm AdS}_3 \times {\mathbb R}$.


\section{Concluding comments}

There are five types of maximally supersymmetric backgrounds 
in four-dimensional  $\cN=1$ off-shell supergravity, two of which are well known: 
Minkowski superspace ${\mathbb R}^{4|4}$ \cite{AV,SS} and 
anti-de Sitter superspace ${\rm AdS}^{4|4}$
\cite{Keck,Zumino77,IS}.
The remaining three  superspaces, 
${\mathbb M}^{4|4}_{T}$, ${\mathbb M}^{4|4}_{S}$ and ${\mathbb M}^{4|4}_{N}$, 
are described by the geometry \eqref{RS^3} 
with different choices of $G_a$. 
All five $\cN=1$ superspaces possess $\cN=2$ extensions. 
The Maxwell-Goldstone multiplet on ${\mathbb R}^{4|4}$
for partially broken $\cN=2$ Poincar\'e supersymmetry 
was found long ago \cite{BG,RT}. 
In this paper,  we have constructed the Maxwell-Goldstone multiplets 
which are defined on ${\mathbb M}^{4|4}_{T}$, ${\mathbb M}^{4|4}_{S}$ 
and ${\mathbb M}^{4|4}_{N}$
and describe partial $\cN=2 \to \cN=1$ supersymmetry breaking.

In Appendix \ref{AppendixC} we demonstrate
that no Maxwell-Goldstone multiplet {\it action}
for partial $\cN=2 \to \cN=1$ supersymmetry breaking 
exists in the case of the anti-de Sitter (AdS) supersymmetry.
The reason for this obstruction is the fact that every covariantly 
constant $\sSU(2)$ triplet $G^{ij}_+$ must be proportional to 
the torsion tensor $S^{ij}$, which is real and  covariantly constant 
in ${\rm AdS}^{4|8}$ \cite{KLRT-M1}. As a consequence, 
the conditions  \eqref{3.2} and \eqref{3.3} are not compatible in 
 ${\rm AdS}^{4|8}$. Since the $\cN=1$ AdS superspace 
 ${\rm AdS^{4|4} } $ is naturally embedded in ${\rm AdS^{4|8} } $ as a subspace
 \cite{KT-M08_conf-flat}, applying the formalism of section \ref{section2}
to the case of ${\rm AdS}^{4|8}  $ allows us to derive  a 
Maxwell-Goldstone multiplet for partially broken $\cN=2$ AdS supersymmetry.
The corresponding technical details are spelled out in Appendix \ref{AppendixC}.
However, since the conditions  \eqref{3.2} and \eqref{3.3} are not compatible in 
 ${\rm AdS}^{4|8}$, we cannot use this Maxwell-Goldstone multiplet to construct 
 a supersymmetric invariant action.

There exists a one-parameter family of $\cN=1$ supersymmetric extensions
of the Born-Infeld actions \cite{CF}. A unique extension is fixed by the requirement 
that the action should describe  
the Maxwell-Goldstone multiplet on ${\mathbb R}^{4|4}$
for partially broken $\cN=2$ Poincar\'e supersymmetry \cite{BG,RT}. 
The same extension is uniquely fixed by the requirement of $\sU(1)$
duality invariance \cite{KT2,KT3}, which implies the self-duality 
under superfield Legendre transform discovered by Bagger and Galperin \cite{BG}.
A curved-superspace extension of the $\cN=1$ supersymmetric Born-Infeld action
is not unique. However, a unique extension is fixed by the requirement of $\sU(1)$
duality invariance \cite{KMcC}. It is given by the action \eqref{curvedAc}
in which $X$ is a unique solution to the constraint 
\bea
X + \frac{1}{4}   X(\bar \cD^2 -4R) 
{\bar X}  = W^2~,
\label{eq:constraint}
\eea
with $R$ the chiral scalar torsion superfield. This action was first proposed 
in \cite{GK}. In the case of anti-de Sitter superspace ${\rm AdS}^{4|4}$, 
the only non-zero components of the superspace torsion are $R$ and $\bar R$, 
which are constant. The corresponding $\cN=1$ supersymmetric Born-Infeld action 
possesses  $\sU(1)$ duality invariance, however it is not invariant under a second 
nonlinearly realised supersymmetry, as demonstrated in Appendix C.
Therefore, this action is not suitable to describe a partial breaking 
of the $\cN=2$ AdS supersymmetry. 

In addition to the Maxwell-Goldstone multiplet of \cite{BG,RT}, 
there exist other multiplets 
for partially broken $\cN=2$ Poincar\'e supersymmetry \cite{BG2,RT,G-RPR}. 
We believe these models can be generalised to the superspaces
 ${\mathbb M}^{4|4}_{T}$, ${\mathbb M}^{4|4}_{S}$ 
and ${\mathbb M}^{4|4}_{N}$
to describe partial $\cN=2 \to \cN=1$ supersymmetry breaking. 
It would also be interesting to investigate whether some of these models
can be extended to describe
partially broken  $\cN=2$ AdS supersymmetry.

Recently, there has been much interest in 
models for  spontaneously broken local 
$\cN=1$ supersymmetry \cite{ADFS,DFKS,BFKVP,HY,AM,Kallosh:2015sea,Kallosh:2015tea,Schillo:2015ssx,Kallosh:2015pho}, which are 
based on the use of the nilpotent chiral Goldstino superfield proposed in \cite{Casalbuoni1989,KSei}.
Other nilpotent Goldstino superfields can be used to describe 
spontaneously broken $\cN=1$ supergravity \cite{LR,KTyler,K15}
(for an alternative approach to de Sitter supergravity, see
\cite{Bandos:2015xnf}).
At the moment it is not clear whether the nilpotent $\cN=2$ chiral superfield
advocated in the present paper is suitable for the description of 
partial supersymmetry breaking in $\cN=2$ supergravity. 
It is certainly of interest to develop a superspace description for the models for spontaneous $\cN=2 \to \cN=1$ local supersymmetry breaking 
pioneered in \cite{FGP,Fre} and further developed, e.g., in \cite{Louis1,Louis2}.

\newpage
\noindent
{\bf Acknowledgements:}\\
We thank Joseph Novak for comments on the manuscript and suggestions.
GT-M is grateful to Daniel Butter for discussions.
This work  is supported
in part by the Australian Research Council (ARC) Discovery Project DP140103925.
The work of GT-M is also supported by the Interuniversity Attraction Poles Programme
initiated by the Belgian Science Policy (P7/37) and in part by COST Action MP1210 
``The String Theory Universe.''


\appendix 

\section{$\cN=1$ superspaces over $\sU(2)=(S^1 \times S^3)/{\mathbb Z}_2$
and $({\rm AdS}_3 \times S^1)/{\mathbb Z}_2$}

In this appendix we give supermatrix realisations for  
two maximally supersymmetric backgrounds in 4D $\cN=1$ supergravity.

\subsection{$\cN=1$ superspace over $\sU(2)=(S^1 \times S^3)/{\mathbb Z}_2$} \label{subsectionA.1}

Here and in the next appendix, the supergroup $\sSU(2|1)$ is defined  
to consist of complex $(2|1) \times (2|1)$  supermatrices
(with $A, D$ bosonic blocks and $B, C$ fermionic ones)
\bea
g = \left(
\begin{array}{c|c}
 A  & B\\
 \hline
C &    D
\end{array}
\right)~
\eea
constrained by
\bea
g^\dagger \eta g &=& \eta~, 
\qquad {\rm Ber} \,g =1~, \qquad 
\eta = \left(
\begin{array}{c | r}
 {\mathbbm  1}_2   ~& 0 \\
 \hline
0 &     -1
\end{array}
\right)~.
\eea

We introduce a superspace $\cM^{4|4}$ consisting of 
complex $(2|1) \times (2|0)$  supermatrices
(with $h$ bosonic  and $\Q$ fermionic blocks)
\bea
\cP= \left(
\begin{array}{c}
 h \\  \Q
\end{array}
\right) 
\label{A.3}
\eea
constrained by 
\bea
\cP^\dagger \eta \cP = {\mathbbm 1}_2 \quad \Longleftrightarrow \quad
h^\dagger h = {\mathbbm 1}_2 + \Q^\dagger \Q~.
\eea
The supermanifold  defined by this equation coincides with the 4D $\cN=1$ 
compactified Minkowski superspace 
(described in detail in  section 3 of \cite{K-compactified}) on which the superconformal 
group $\sSU(2,2|1)$ acts by well-defined transformations. 
The bosonic body of the superspace is $\sU(2)=(S^1 \times S^3)/{\mathbb Z}_2$. 

It is useful to switch from the variables $h$ and $\Q$ to new ones, 
$\vf \in \mathbb R$, $u$ and $\q$,   
defined as follows:
\bea
\cP= \left(
\begin{array}{c}
\re^{\ri \vf }  u \\  
\e^{\ri \vf} \q
\end{array}
\right) ~, \qquad u^\dagger u = {\mathbbm 1}_2 + \q^\dagger \q~, 
\qquad \det u = \det u^\dagger = \big( 1 + \q \q^\dagger \big)^{-\hf}~. 
\label{A.5}
\eea
We can represent 
\bea
u = \hat u \sqrt{ {\mathbbm 1}_2 + \q^\dagger \q}~, 
\qquad \hat u \in \sSU(2)~.
\label{A.6}
\eea
The supermatrix \eqref{A.5} is invariant under the ${\mathbb Z}_2$ transformation 
$\vf \to \vf +\p$, $\hat u \to -\hat u$ and $\q \to -\q$.
This is the origin of ${\mathbb Z}_2$ in $\sU(2)=(S^1 \times S^3)/{\mathbb Z}_2$. 

It turns out that the superspace $\cM^{4|4}$ introduced above can be identified with 
the group manifold $\sSU(2|1)$. Indeed, it may be checked that 
every element $g \in \sSU(2|1)$ has the form (compare with a similar result in \cite{KS})
\bea
g = \left(
\begin{array}{c|c}
 \re^{\ri \vf }  u   ~& ~\re^{2\ri \vf} \big( 1 + \q \q^\dagger \big)^{-\hf} u \q^\dagger \\
 \hline
\e^{\ri \vf} \q ~&    \re^{2\ri \vf} (1 + \q \q^\dagger)^\hf
\end{array}
\right)~,
\label{A.7}
\eea
where $u$ is constrained as in \eqref{A.5}. 

The isometry group of  $\cM^{4|4}$ is $\sSU(2|1) \times \sU(2) $.
It acts on  $\cM^{4|4}$ as follows:
\bea
\cP ~\to ~ g_{\rm L} \cP g_{\rm R}^{-1} ~, \qquad
g_{\rm L} \in \sSU(2|1) ~, \qquad g_{\rm R} \in \sU(2)~.
\eea 
These transformations are holomorphic in terms of the variables 
$h$ and $\Q$ (hence the isometry transformations act on a chiral subspace 
of the full superspace).
The isometry group has two $\sU(1)$ subgroups that describe $R$-symmetry 
transformations and time translations. One subgroup corresponds 
to all diagonal supermatrices \eqref{A.7} with $u={\mathbbm 1}_2$ and $\q=0$. 
The other subgroup is spanned by all diagonal matrices $\re^{\ri \j}{\mathbbm 1}_2$
in $\sU(2)$.

On the group manifold $\sSU(2|1)$, we can define an action 
of $\sSU(2|1) \times \sSU(2|1)$ by the standard rule
\bea
g ~\to g_{\rm L} \,g \,g_{\rm R}^{-1} ~, \qquad
g_{\rm L}, g_{\rm R} \in \sSU(2|1) ~.
\eea
These transformations leave invariant the supermetric 
\bea
\rd s^2 = -\hf {\rm Str} \, \cE^2~, \qquad \cE = g^{-1} \rd g~.
\eea
However, such transformations map the chiral subspace \eqref{A.3} to itself only if 
$g_{\rm R} \in \sU(2)$.


\subsection{$\cN=1$ superspace over 
$\sU(1,1)=({\rm AdS}_3 \times S^1)/{\mathbb Z}_2$} \label{subsectionA.2}

We define the supergroup $\sSU(1,1|1)$   
to consist of complex $(2|1) \times (2|1)$  supermatrices
(with $A, D$ bosonic blocks and $B, C$ fermionic ones)
\bea
g = \left(
\begin{array}{c|c}
 A  & B\\
 \hline
C &    D
\end{array}
\right)~
\eea
constrained by
\bea
g^\dagger \eta g &=& \eta~, 
\qquad {\rm Ber} \,g =1~, \qquad 
\eta = \left(
\begin{array}{c | r}
 \s_3   ~& 0 \\
 \hline
0 &     -1
\end{array}
\right)~.
\eea
Every element $g \in \sSU(1,1|1)$ can be written in the form
\bea
g = \left(
\begin{array}{c|c}
 \re^{\ri \vf }  u   ~& ~\re^{2\ri \vf} \big( 1 + \q \s_3\q^\dagger \big)^{-\hf} u \q^\dagger \\
 \hline
\e^{\ri \vf} \q ~&    \re^{2\ri \vf} (1 + \q \s_3\q^\dagger)^\hf
\end{array}
\right)~,
\label{A.13}
\eea
where $u$ is constrained by 
\bea
 u^\dagger \s_3 u = \s_3 + \q^\dagger \q~, 
\qquad \det u = \det u^\dagger = \big( 1 + \q \s_3\q^\dagger \big)^{-\hf}~. 
\label{A.14}
\eea
We can represent 
\bea
u = \hat u \sqrt{ {\mathbbm 1}_2 + \s_3 \q^\dagger \q}~, 
\qquad \hat u \in \sSU(1,1)~.
\label{A.15}
\eea
The supermatrix defined by eqs. \eqref{A.13} and \eqref{A.15}
is invariant under the discrete  transformation 
$\vf \to \vf +\p$, $\hat u \to -\hat u$ and $\q \to -\q$.

We introduce a four-dimensional superspace $\widetilde{\cM}^{4|4}$ consisting of 
complex $(2|1) \times (2|0)$  supermatrices
(with $h$ and $\Q$ being bosonic  and fermionic blocks, respectively)
\bea
\cP= \left(
\begin{array}{c}
 h \\  \Q
\end{array}
\right) \equiv
 \left(
\begin{array}{c}
 \re^{\ri \vf }  u \\   \re^{\ri \vf }  \q
\end{array}
\right) 
 ~,
\label{A.17}
\eea
where $\vf$, $u$ and $\q$ are defined as in \eqref{A.13}.
This superspace can be identified with  the group manifold $\sSU(1,1|1)$.
Its bosonic body is $\sU(1,1)=({\rm AdS}_3 \times S^1)/{\mathbb Z}_2$. 

The isometry group of $\widetilde{\cM}^{4|4}$ is $\sSU(1,1|1) \times \sU(2) $.
It acts on $\widetilde{\cM}^{4|4}$ as follows:
\bea
\cP ~\to ~ g_{\rm L} \cP g_{\rm R}^{-1} ~, \qquad
g_{\rm L} \in \sSU(1,1|1) ~, \qquad g_{\rm R} \in \sU(2)~.
\eea 
These transformations are holomorphic in terms of the variables 
$h$ and $\Q$ (hence the isometry transformations acts on a chiral subspace 
of the full superspace), and leave invariant the supermetric 
\bea
\rd s^2 = \hf {\rm Str} \, \cE^2~, \qquad \cE = g^{-1} \rd g~.
\eea
Unlike the superspace considered in the previous subsection, the dimension
parametrised by $\vf$ is now space-like.

Let us consider the coset space
\bea
{\rm AdS}_{(3|2,0)} :=  \sSU(1,1|1) /\sU(1)~,
\eea
where the subgroup $\sU(1)$ of $ \sSU(1,1|1)$ consists of 
all diagonal supermatrices \eqref{A.13}
with $u = {\mathbbm 1}_2$ and $\q=0$. 
This coset space
may be seen to coincide with the 3D (2,0) anti-de Sitter superspace \cite{KT-M11}. 
We recall that in three dimensions, $\cN$-extended anti-de Sitter (AdS) superspace exists
in several incarnations known as $(p,q)$ AdS superspaces, where
the  non-negative integers $p \geq q$ are such that $\cN=p+q$.   
The conformally flat $(p,q)$ AdS superspace is
\bea
{\rm AdS}_{(3|p,q)} = \frac{ {\sOSp} (p|2; {\mathbb R} ) \times  {\sOSp} (q|2; {\mathbb R} ) } 
{ {\sSL}( 2, {\mathbb R}) \times {\sSO}(p) \times {\sSO}(q)}~.
\eea
In the case $p=\cN\geq 4$ and $q=0$, non-conformally flat AdS superspaces
also exist \cite{KLT-M12}. 


\section{$\cN=2$ superspace over  $\sU(2)=(S^1 \times S^3)/{\mathbb Z}_2$}

$\cN=2$ superspace $\cM^{4|8}$ over  $\sU(2)=(S^1 \times S^3)/{\mathbb Z}_2$
can be realised as the quotient space
\bea
\cM^{4|8} := \cM^{4|4}_{\rm L} \times \cM^{4|4}_{\rm R} \Big/ \sim~,
\eea
where $\cM^{4|4}_{\rm L}$ and  $\cM^{4|4}_{\rm R} $ denote two copies 
of $\cM^{4|4}$. The equivalence relation is defined by the rule: 
two pairs ${\bm \cP} =(\cP_{\rm L} , \cP_{\rm R})$ and 
${\bm \cP}' = (\cP'_{\rm L} , \cP'_{\rm R} )$ 
are equivalent, ${\bm \cP} \sim {\bm \cP}'$,  if 
\bea
\cP'_{\rm L} = \cP_{\rm L} h~, \qquad \cP'_{\rm R} = \cP_{\rm R} h~, 
\eea
for some group element $h \in \sU(2)$. 

The isometry group of $\cM^{4|8}$ is 
\bea
{\bm G} := G_{\rm L} \times G_{\rm R} \times \sU(1)
= \sSU(2|1) \times \sSU(2|1) \times \sU(1)~.
\eea
Given a group element ${\bm g} = g_{\rm L} \times g_{\rm R} \times \re^{\ri \j} \in 
{\bm G}
$, with $\j \in {\mathbb R}$, 
it acts on the pair ${\bm \cP} =(\cP_{\rm L} , \cP_{\rm R})$ by the rule: 
\bea
(\cP_{\rm L} , \cP_{\rm R})  ~\to ~ ({\bm g} \cP_{\rm L} , {\bm g} \cP_{\rm R})~, 
\qquad 
{\bm g} \cP_{\rm L} = g_{\rm L} \cP_{\rm L} \re^{\ri \j}
~, \qquad
{\bm g} \cP_{\rm R} = g_{\rm R} \cP_{\rm R} \re^{-\ri \j}~.
\eea

The equivalence relation allows us to choose $\cP_{\rm R}$ in the form:
\bea
\cP_{\rm R}= \left(
\begin{array}{c}
\sqrt{ {\mathbbm 1}_2 + \j^\dagger \j}~ \\  
 \j
\end{array}
\right) ~.
\eea

The above construction can readily be modified in order to describe 
the  $\cN=2$ superspace over $\sU(1,1)=({\rm AdS}_3 \times S^1)/{\mathbb Z}_2$.


\section{Example: The anti-de Sitter supersymmetry}
\label{AppendixC}

In this appendix we show that the formalism of sections 2 and 3 can be used
to define a Goldstone-Maxwell multiplet for partially broken 4D $\cN=2$
anti-de Sitter (AdS) supersymmetry with the following properties: 
(i) it is the standard Maxwell multiplet with respect to the $\cN=1$ AdS supersymmetry; 
(ii) it transforms nonlinearly under the second AdS supersymmetry. 
However, making use of this multiplet does 
not allow one to construct an invariant action describing the partial 
$\cN=2 \to \cN=1$ AdS supersymmetry breaking.

To start with, we recall a few definitions concerning 
the 4D $\cN=2$ AdS superspace 
$$
{\rm AdS^{4|8} } := \frac{{\sOSp}(2|4)}{{\sSO}(3,1) \times {\sSO} (2)}~,
$$
which is a maximally symmetric geometry that originates within 
the off-shell formulation for $\cN=2$ conformal supergravity developed in  \cite{KLRT-M1}. For comprehensive studies of $\cN=2$ supersymmetric field theories
in AdS${}_4$, the reader is referred to \cite{KT-M08_conf-flat,BKLT-M}. 

We assume that ${\rm AdS^{4|8} }$  is parametrised  by local bosonic ($x$) and fermionic ($\q, \bar \q$) 
coordinates  ${\bm z}^{\cM}=(x^{m},\q^{\mu}_{\imath},{\bar \q}_{\dot{\mu}}^{\imath})$
(where $m=0,1,2,3$, $\mu=1,2$, $\dot{\mu}=1,2$ and  $\imath=\1,\2$).
The corresponding covariant derivatives
\bea
{\cD}_{\cA} =({ \cD}_{a}, { \cD}_{{\a}}^i, { \cDB}^\ad_i)
= E_{\cA}{}^\cM \pa_\cM + \hf  {\O}_{\cA}{}^{ bc} M_{ bc} 
+  \F_{\cA}{}^{ij} J_{ij}~, \qquad i,j =\1 , \2
\label{1.1}
\eea
obey the algebra  \cite{KT-M08_conf-flat}
\begin{subequations}\label{1.222}
\bea
&\{\cD_\a^i,\cD_\b^j\}=
4{S}^{ij}M_{\a\b}
+2 \ve_{\a\b}\ve^{ij}S^{kl}J_{kl}~,
\qquad
\{\cD_\a^i,\cDB^\bd_j\}=
-2\ri\d^i_j(\s^c)_\a{}^\bd\cD_c
~,~~~
\\
&{[}\cD_a,\cD_\b^j{]}=
\frac{\ri}{2} ({\s}_a)_{\b\gd}S^{jk}\cDB^\gd_k~, \qquad
{[}\cD_a,\cDB^\bd_j{]}=
\frac{\ri}{ 2} (\tilde{\s}_a)^{\bd\g}S_{jk}\cD_\g^k
~,~~~~ \\
&
[\cD_a,\cD_b]= - S^2
M_{ab}~.
~~~~~~
\label{AdS-N2-2}
\eea
\end{subequations} 
The $\sSU(2)$ triplet  ${S}^{ij} $ is the only non-vanishing component of the superspace torsion in 
${\rm AdS^{4|8} }$;
it is {\it covariantly constant} and real
 \bea
 \cD_\cA S^{ij}=0~, \qquad \bar S^{ij} = S^{ij}~.
 \eea
The parameter ${S}^2 :=  \frac{1}{2} { S}^{ij} { S}_{ij} = \text{const}$ 
is positive, and therefore \eqref{AdS-N2-2} gives the
algebra of covariant derivatives in AdS${}_4$.

The isometry transformations of AdS$^{4|8}$ form the supergroup  $\sOSp(2|4)$. 
In the infinitesimal case,  an isometry transformation 
is described by a Killing  supervector field 
$\x^{\cA} E_\cA$, with $E_\cA= E_{\cA}{}^\cM \pa_\cM$, 
defined to obey the equation
\bea
\big[\x+\hf l^{bc}M_{bc}
+\r S^{jk}J_{jk}
,\cD_{\cA}\big] =0~, \qquad
\x:=\x^{\cB}  \cD_{\cB}
= \x^b\cD_b+\x^\b_j\cD_\b^j + \xb_\bd^i\cDB^\bd_j~,
\label{Super-K-eq}
\eea
for some real antisymmetric tensor $l^{bc}(z)$ and scalar $\r(z)$ parameters.
It turns out that the Killing equation (\ref{Super-K-eq}) uniquely determines 
the parameters $\x^\a_i$, $l^{cd}$ and $\r$ in terms of $\x^a$. A similar property exists
for superspace isometry transformations in any number of dimensions \cite{K15Corfu}.
The specific feature of the 4D $\cN=2$ AdS superspace is that the parameters
$\x^\cA$ and $l^{ab}$ are uniquely expressed in terms of $\r$  \cite{KT-M08_conf-flat}.

Due to \eqref{1.222}, the $\sSU(2)$ gauge freedom can be used to choose the $\sSU(2)$ connection
$\F_{\cA}{}^{ij} $ in 
\eqref{1.1}
to look like $ \F_{\cA}{}^{ i j } =  \F_{\cA} S^{ij}$, for some one-form $\F_\cA$
describing the residual $\sU(1)$ connection associated with the generator $S^{ij} J_{ij}$.  
Then $S^{ij}$ becomes a constant iso-triplet, $S^{ij} =  {\rm const}$.
The remaining global $\sSU(2)$ rotations
can take $S^{ij}$ to any position on the two-sphere of radius 
$S$.
We make the choice
\be
{S}^{\1 \2} =0~,  \qquad \m:= -S^{\2\2}~, \qquad
\mub=-S^{\1\1}  ~,  
\label{S11-mu}
\ee
with $|\m| = S$.
This choice
 must be used in order to embed an $\cN=1$ AdS superspace,
${\rm AdS^{4|4} }$, into the full $\cN=2$ AdS superspace 
\cite{KT-M08_conf-flat}.

As already mentioned, the choice ${S}^{\1 \2} =0$ is required
for embedding ${\rm AdS^{4|4} }$ into ${\rm AdS^{4|8} }$. 
By applying certain general coordinate and local $\sU(1)$ transformations in ${\rm AdS^{4|8} }$, 
it is possible to identify  ${\rm AdS^{4|4} }$ with the surface $\q^{\mu}_{\2} = 0 $ and 
${\bar \q}_{\dot{\mu}}^{\2} =0$. The covariant derivatives for ${\rm AdS^{4|4} }$,
\be
\cD_{A}=(\cD_a,\cD_\a,{\bar \cD}^\ad)
= E_A{}^M\pa_M+\hf\O_A{}^{bc}M_{bc}~,
\ee 
are related to \eqref{1.1} as follows 
\be
\cD_\a := \cD_\a^{\1}\big|~, \qquad
\bar \cD^\ad := \bar \cD^\ad_{\1} \big|~,
\ee
and similarly for the vector covariant derivative.
Here the 
bar-projection is defined by 
\be
U | :=  U(x,\q_{\imath},\bar \q^{\imath})|_{\q_\2={\bar \q}^\2=0}~,
\label{bar-projection}
\ee 
for any $\cN=2$  tensor superfield  $ U(x,\q_{\imath},\bar \q^{\imath})$. It follows from \eqref{1.222} 
that  the $\cN=1$ covariant derivatives obey the algebra 
\bsubeq \label{1.5}
\bea
&\{\cD_\a,\cD_\b\}=-4\bar{\mu}M_{\a\b}~,~~~
\{\cDB_\ad,\cDB_\bd\}=4\mu\bar{M}_{\ad\bd}~,~~~
\{\cD_\a,\cDB_\bd\}=-2\ri\cD_{\a\bd}
~,
\\
&{[}\cD_a,\cD_\b{]}=-\frac{\ri}{2}\bar{\mu}(\s_a)_{\b\gd}\cDB^{\gd}~,~~~
{[}\cD_a,\cDB_\bd{]}=\frac{\ri}{2}\mu(\s_a)_{\g\bd}\cD^{\g}~, \\
&{[}\cD_a,\cD_b{]}=-|\mu|^2 M_{ab}
~,
\eea
\esubeq
which indeed corresponds to the $\cN=1$ AdS superspace (see \cite{BK} for more details).
As a result, every $\cN=2$ supersymmetric field theory in 
${\rm AdS^{4|8} }$ can be reformulated as some theory in ${\rm AdS^{4|4} }$. 

Given an $\cN=2$  tensor superfield  $ U(x,\q_{\imath},\bar \q^{\imath})$,
its infinitesimal  $\sOSp(2|4)$ transformation law is 
\bea
\d_\x U = -\big(\x+\hf l^{bc}M_{bc} +\r S^{jk}J_{jk}\big) U~.
\label{C.100}
\eea
Upon reduction to ${\rm AdS^{4|4} }$, this transformation law turns into a superposition of
several independent $\cN=1$ transformations.
Evaluating the bar-projection of $\x$ gives
\bsubeq
\bea
&\x|=\l +\ve^\a\cD_\a^\2|+\bar{\ve}_\ad\cDB^\ad_\2|~,\qquad
\l =\l^A \cD_A= \l^a\cD_a+\l^\a\cD_\a+\lb_\ad\bar \cD^\ad~,
\eea
where we have introduced 
\bea
&\l^{a}:=\x^{a}|~, \quad 
\l^\a:=\x^\a_\1|~,\quad 
\lb_\ad:=\xb_\ad^\1|~,\qquad 
\ve^\a:=\x^\a_\2|~, \quad \bar{\ve}_\ad:=\x_\ad^\2|
~.
\eea
\esubeq
We denote the bar-projection of the parameters $l_{ab} $ and $\r$ as
\bea
&&\o_{ab}:=l_{ab}|~,\qquad
\ve:=\r|~.
\eea
It holds that 
\bea
\o_{\a\b} = \cD_\a\l_\b= \cD_\b\l_\a~.
\eea
Now, the bar-projection of \eqref{C.100} takes the form
\bea
\d_\x U|&=&-\Big(\l+\hf\o^{ab}M_{ab} \Big) U| 
-\Big( \ve^\a( \cD_\a^\2 U)| 
+\bar{\ve}_\ad (\cDB^\ad_\2 U)|\Big) 
~+~\ve(
{\bar \m}J_{\1\1}+
\m
J_{\2\2})U|~.
~~~~~~
\label{deltaU|}
\eea
The first term on the right is an infinitesimal  $\sOSp(1|4)$
transformation generated by $\l$. The parameters $\l$ and $\o^{bc}$ obey the equation
\bea
{[}\l+ \hf \o^{bc}M_{bc},\cD_A{]}=0~,
\label{N=1-killings-0}
\eea
which defines the  Killing supervector field of AdS$^{4|4}$ \cite{BK}.
The second and third terms on the right of \eqref{deltaU|}
prove to describe the second supersymmetry and $\sU(1)$ transformations.
The corresponding parameters $\ve_\a,\,\bar{\ve}_\ad$ and $\ve$
have the properties 
\bea
&&\ve_\a=\hf \cD_{\a}\ve~, \qquad
\cD_\a\bar \cD_\ad\ve=0~, \qquad 
\big(\cD^2-4\bar{\mu}\big)\ve=0~.
\label{epsilon}
\eea
The parameter $\ve$ was originally introduced in \cite{GKS}.

We are now prepared to analyse the nilpotent $\cN=2$ chiral superfield $\cZ$
constrained by \eqref{Z} in the case that the background superspace is AdS$^{4|8}$.
We recall that a necessary ingredient of the construction described in section 3 is 
that $G^{ij}$ is covariantly constant, $\cD_\cA G^{ij}=0$.
We require this condition to hold in AdS$^{4|8}$, 
which implies that $G^{ij}$ is proportional to $S^{ij}$
\bea
G^{ij}=\k S^{ij}
~,
\eea
where  $\k$ is a real constant.
In accordance with \eqref{S11-mu}, we have $G^{\1\2}=0$.
The parameter $\k$ can be chosen to have any given non-zero value by means 
of rescaling the chiral superfield $\cZ$. We choose $\k =|\m|$, 
and hence
$G^{\1\1}=-|\m| \bar{\mu}$ and $G^{\2\2}=-|\m|\mu$.

The degrees of freedom described by $\cZ$ 
are those of an Abelian $\cN=1$ vector multiplet in  AdS$^{4|4}$.
Indeed, upon reduction to the $\cN=1$ AdS superspace, the $\cN=2$ chiral 
scalar $\cZ$  leads to two chiral superfields, $X$ and $W_\a$, defined as
\begin{subequations}
\bea
X&:=&\cZ|
~,\qquad \bar \cD_\ad X=0~,\\
W_\a
&:=&
-\frac{\ri}{2}\cD_\a^\2\cZ|
~,\qquad
\bar \cD_\ad W_\a=0
~.
\eea
\end{subequations}
One may check that 
the $\cN=2$ constraints \eqref{Z} imply the Bianchi identity
\bsubeq\label{constr-X-1111112}
\bea
\cD^\a W_\a&=&\bar \cD_\ad \bar{W}^{\ad}
~,
\label{C.19a}
\eea
as well as the nonlinear  constraint
\bea
-\ri\mu |\m|X
+\frac{1}{4}X\big(
\bar \cD^2
- 4\mu
\big)\bar{X}
&=&
W^2
~,~~~~~~
W^2:=W^\a W_\a
~.
\label{constr-X-111111}
\eea
\esubeq
Eq. \eqref{C.19a} tells us that $W_\a$ is the chiral field strength of a Maxwell multiplet
in AdS$^{4|4}$.
Eq. \eqref{constr-X-111111} is of the same type as the constraint \eqref{eq:constraint},
which generates the $\cN=1$ locally supersymmetric Born-Infeld action 
with $\sU(1)$ duality invariance.
The constraint  \eqref{constr-X-111111} is uniquely solved by 
expressing $X$ in terms of $W^2$ and $\bar{W}^2$ 
and their covariant derivatives,
in complete analogy with the general supergravity analysis of \cite{KMcC}.

In accordance with \eqref{C.100}, the  infinitesimal  $\sOSp(2|4)$ transformation of $\cZ$ is 
$\d \cZ = -\x \cZ$. Using this result,  
it is straightforward to derive the transformation laws of $X$ and $W_\a$
under the second supersymmetry and $\sU(1)$ transformations described by the superfield parameter
$\ve$. Making use of the constraints obeyed by $\cZ$ and $X$,
we obtain
\begin{subequations}\label{C.200}
\bea
\d_\ve X&=&
-2\ri\ve^\a W_\a
~, \\
\d_\ve W_\a&=&
\ri\ve_\a\Big{[}\,
\ri\mu |\m|
-\frac{1}{4}\big(\bar \cD^2- 4\mu\big)\bar{X} 
-\mu X
\Big{]}
-\bar{\ve}^\bd \cD_{\a\bd} X
+\frac{\ri}{2}\m\ve  \cD_{\a}X
~.
\eea
One can check that $\d_\ve X$ and $\d_\ve W_\a$ 
preserve the constraints \eqref{constr-X-1111112}. Due to
\eqref{epsilon}, the variation $\d_\ve W_\a$ can be rewritten in the form
\bea
\d_\ve W_\a&=&
-\frac{\ri}{8}\big(\bar \cD^2- 4\mu\big)\Big{[}
2\Big(
\bar{X} 
-X
+\ri |\m|
\Big)\ve_\a
-\ve \cD_\a X
\Big{]}
~,
\eea
\end{subequations}
which 
makes manifest the chirality  of $\d_\ve W_\a $. It follows from \eqref{C.200}
that the second supersymmetry and $\sU(1)$ transformations are nonlinearly realised.

Let us consider  the  supersymmetric and $\sU(1)$ duality invariant  Born-Infeld action in the
$\cN=1$ AdS superspace\footnote{In accordance with \eqref{constr-X-111111}, 
the overall coefficient in \eqref{C.222} is chosen such that the kinetic term for the vector multiplet 
is canonically normalised, 
$S=\frac{1}{4}
\int {\rm d}^4x \,{\rm d}^2 \q \, \cE\,W^2 
+{\rm c.c.} +\mbox{interaction terms}$.
It should be remarked that the functional ${\rm Re} \big(\m
\int {\rm d}^4x \,{\rm d}^2 \q \, \cE\,X \big)$ is a total derivative, 
in accordance with  \eqref{constr-X-111111}. }
\bea
S=
-\frac{\ri}{4}|\m|\m
\int {\rm d}^4x \,{\rm d}^2 \q \, \cE\,X
+{\rm c.c.}~,
\label{C.222}
\eea
with $X$ constrained by \eqref{constr-X-111111}.
The action is manifestly invariant under the isometry transformations of AdS$^{4|4}$, 
with the infinitesimal transformation law of $W_\a$ being 
\bea
\d W_\a = -\l W_\a -\o_\a{}^\b W_\b~.
\eea
However, the action is not invariant under the transformation  \eqref{C.200}, 
\bea
\d_\ve
 S&=& 2|\m|^3
\int \rd^4 x \, {\rm d}^2\q \rd^2 \bar \q \, E\,\ve V
~,
\label{C.23}
\eea
where the real scalar $V$ denotes the unconstrained prepotential of the  vector multiplet, 
\bea
W_\a=-\frac{1}{4}(\bar \cD^2-4\mu)\cD_\a V
~.
\eea
Eq. \eqref{C.23}
is a unique feature that distinguishes AdS${}_4$ from the other maximally supersymmetric backgrounds 
we have studied in this paper.


\begin{footnotesize}

\end{footnotesize}

\end{document}